\documentclass[pre,showpacs,twocolumn,floatfix,amsmath,amsfonts,longbibliography]{revtex4}
\usepackage{graphicx,epstopdf,url}
\usepackage{color}
\usepackage[colorlinks=true,urlcolor=blue,citecolor=blue,linkcolor=blue,urlcolor=blue]{hyperref}
\usepackage[active]{srcltx}

\begin{document}

\title{
Wrinkle formation during uniaxial compression of a graphene sheet lying on a soft (polymer) substrate
}

\author{Alexander V. Savin$^{1,2}$}
\email{asavin00@gmail.com}
\affiliation{
$^1$N.N. Semenov Federal Research Center for Chemical Physics of the Russian Academy of Sciences, 4 Kosygin St., Moscow 119991, Russia \\
$^2$Plekhanov Russian University of Economics, 36 Stremyanny Lane, Moscow 117997, Russia 
}

\begin{abstract}
Modeling of wrinkles and folds formation in single and multilayer graphene sheets lying on flat deformable (polymer) substrates has been carried out. 
It is shown that the deformability of the substrate leads to the appearance of significant features. In contrast to the flat surfaces of rigid crystals molecules of soft polymer substrates can penetrate into wrinkles and folds of the graphene sheet, completely filling the voids beneath the sheet.
Moreover, the vertical folds of the sheet can be directed not only upward from the substrate, but also down into the substrate, penetrating it.
By modeling the uniaxial compression of the two-component graphene/polymer system, the external pressure and thermal vibrations of the substrate molecules have been taken into account. 
High external pressure $p\ge 150$~bar leads to a noticeable additional stabilization of the initial ground state of the system.
At uniaxial compression above the critical value, a system of localized wrinkles whose interior is filled with molecules of the substrate appears in the graphene sheet.
Increasing temperature leads to an increase in the size of wrinkles and to a decrease in their number.
The largest wrinkles form before the substrate begins to melt.
Melting leads to the complete disappearance of all wrinkles and small folds.
Cooling of the melted substrate leads to its crystallization, but the system of wrinkles in the graphene sheet on the surface is not restored.
Therefore, melting of the polymer substrate and its subsequent cooling can serve as a method getting rid of localized wrinkles and folds of the graphene sheet.
\\ \\
Keywords:
Graphene, graphene wrinkles and folds, polymer substrate, uniaxial compression
\end{abstract}
\maketitle

\section{Introduction}
Carbon atoms are capable of creating numerous structures, out of which graphene (a monoatomic crystalline layer) attract most attention of researchers 
\cite{Novoselov2004,Geim2007,Soldano2010,Baimova2014,Baimova2014a}.
This nanomaterial is of interest because of its unique electronic \cite{Geim2009}, mechanical \cite{Lee2008} and thermal properties \cite{Baladin2008,Liu2015}.
Because of its strength and one-atom thickness, graphene is an ideal candidate for fillers used in polymer nanocomposites \cite{Stankovich2006,Li2012}.
Its exceptional electronic properties allow it to be used in stretchable electronics \cite{Kim2009} if it is placed on soft polymer substrates \cite{Li2016}.

A popular method for producing graphene is chemical vapor deposition (CVD), in which graphene is grown on a substrate in a carbon-rich environment.
The CVD method often results in topological defects (during the cooling process, the graphene sheet undergoes out-of-plane strain bending), such as ripples \cite{Tapaszto2012} and wrinkles \cite{Zhu2012}.
Defects of this type can be formed due to the roughness of the substrate \cite{Lui2009} and due to the different thermal expansion of graphene and substrate \cite{Obraztsov2007}.
The presence of such defects can change the properties of graphene, namely its electrical conductivity \cite{Zhu2012}, thermal conductivity \cite{Chen2012,Wang2014} and elasticity \cite{Wang2011}.
The wrinkle and fold structures arising on the sheet can be used as channels for fluid injection and storage between graphene and its substrate \cite{Carbone2019}, as well as for its spatially selective chemical functionalization \cite{Deng2019}.
Therefore, understanding the laws of wrinkle and fold formation is important for the design of graphene-based nanodevices.

The out-of-plane (transverse) deformations of graphene can be categorized into ripples (corrugations), wrinkles and folds depending on their physical size and topology \cite{Deng2016,Deng2018}.
To describe individual wrinkles and folds, the quasi-analytical models based on calculus of variations \cite{Zhu2012,Zhang2013a,Box2015,Aljedani2020,Aljedani2020a,Cox2020,Aljedani2021}, models based on continuum mechanics using finite element method \cite{Zhang2013,Zhang2014} and full-atom models using molecular dynamics \cite{Mulla2015,Zhu2020,Zhao2020} were used. 
In all these papers, the rigid (non-deformable) substrate approximation was used. In \cite{Li2016}, biaxial compression of a sheet of graphene lying on a deformable polymer substrate was modeled by the molecular dynamics method.
It was shown that the deformability of the substrate significantly affects the morphology of the sheet during its compression.

The molecular dynamics method with full-atomic models currently does not allow to simulate the compression of sufficiently large graphene sheets. 
The size problem can be solved by switching to the models with "united"\ atoms.
Recently, to describe the dynamics of wrinkles and folds of graphene nanoribbons lying on a flat substrate, a two-dimensional chain model describing the longitudinal cross-section of the nanoribbon \cite{Savin2019prb} has been proposed.
The aim of the present work is to explain, using this model, the peculiarities of wrinkle formation
under uniaxial compression in graphene sheets lying on soft polymer substrates. 
The influence of external pressure and thermal fluctuations of the substrate on the processes of wrinkle formation will be evaluated. 
It will be shown that the melting of the substrate leads to the complete disappearance of all wrinkles and small folds.

\section{2D model of multilayer graphene sheet on soft polymer substrate}

\begin{figure}[tb]
\begin{center}
\includegraphics[angle=0, width=0.7\linewidth]{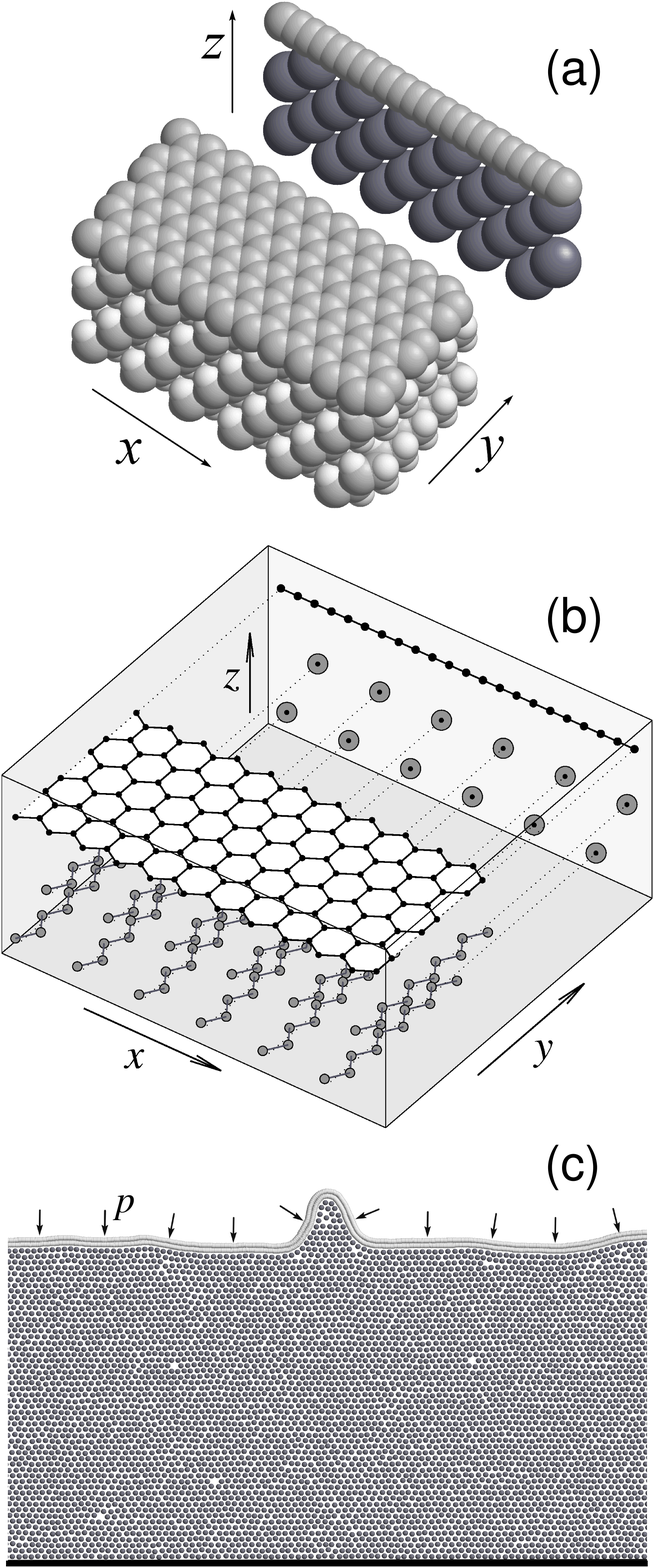}
\end{center}
\caption{\label{fig01}\protect
Construction of a 2D chain model of a graphene nanoribbon lying on a polymer substrate.
(a) Full-atom 3D model of the nanoribbon lying on the flat surface of polyethylene crystal and projection of the molecular system on the $xz$ plane.
(b) A 3D model of the united atoms and its reduction to a 2D model in the $xz$ plane.
(c) 2D model of a bilayer graphene nanoribbon (light disks) lying on a soft polymer substrate (dark disks). 
The arrows show the action of external pressure $p$, the black straight line corresponds to a fixed flat surface on which the polymer substrate lies (a periodic boundary condition is used along the $x$-axis).
}
\end{figure}

For an elastically isotropic graphene sheet, its longitudinal and bending stiffnesses depend weakly on its orientation.
For definiteness, let us consider a graphene nanoribbon whose longitudinal edges have zigzag structure (Fig.~\ref{fig01}). 
The longitudinal and bending transversely isotropic vibrations of the sheet can be described using only the dynamics of the molecular chain, which is a linear cross-section of the sheet. 
Such a 2D model of the chain describing the longitudinal and bending motions of the nanoribbon is presented in Refs. \cite{Savin2015prb,Savin2015ftt}. 
This model was used to describe wrinkles and folds of of graphene sheets on solid flat substrates \cite{Savin2019prb,Savin2024ftt}, and bending deformations of the sheets located in the polymer matrix \cite{Savin2021vms}.

The scheme of 2D chain model construction of a multilayer graphene sheet lying on a soft flat polymer substrate is presented in Fig.  \ref{fig01}.
For a single-layer sheet of graphene lying in the $xy$ plane with a zigzag structure along the $x$ axis the model describes a linear cross-section of the sheet in which one particle corresponds to all atoms of the sheet, having the same coordinates $x$ and $z$. 
In transversely isotropic vibrations, all these atoms move synchronously as one unified atom, changing only the $x$ and $z$ coordinates, but not the $y$ coordinate.
To describe a multilayer sheet of graphene, it is convenient to use a two-dimensional molecular chain model for each layer.

For a single-layer sheet of graphene lying in a plane parallel to the $xy$ plane, with a zigzag structure along the $x$ axis, the model describes a longitudinal cross-section of the sheet in which one particle corresponds to all atoms having the same $x$ coordinate. 
At transversely isotropic oscillations all these atoms move synchronously, changing only coordinates $x$ and $z$, but not changing coordinate $y$. 
In this case, the Hamiltonian of the sheet can be written as Hamiltonian of a 2D chain of $N_c$ links
\begin{eqnarray}
H_1=\sum_{n=1}^{N_c}\left[\frac12M_c(\dot{\bf u}_n,\dot{\bf u}_n)+V({\bf u}_n,{\bf u}_{n+1})
\right.\nonumber\\
\left.+U({\bf u}_{n-1},{\bf u}_n,{\bf u}_{n+1})+\frac12\sum_{|k-n|>5}W_1({\bf u}_n,{\bf u}_{k})\right],
\label{f1}
\end{eqnarray}
where the two-dimensional vector ${\bf u}_n=(x_n,z_n)$ defines the coordinates of the $n$th particle of the chain having mass $M_c=12m_p$ ($m_p=1.66\cdot 10^{-27}$~kg is the proton mass).

The first potential
\begin{eqnarray}
V({\bf u}_n,{\bf u}_{n+1})=\frac12K(r_n-a)^2, \label{f2}\\
r_n=|{\bf v}_n|,~ {\bf v}_n={\bf u}_{n+1}-{\bf u}_n, \nonumber
\end{eqnarray}
defines the interaction between neighboring nodes of the chain, equilibrium distance (chain pitch) $a=1.228$~\AA, longitudinal stiffness $K=405$~N/m ($r_n$ is distance between neighboring links $n$ and $n+1$, $a=r_c\sqrt{3}/2$, $r_c=1.418$~\AA~ is the length of the C--C valence bond in graphene).

The second potential
\begin{eqnarray}
U({\bf u}_{n-1},{\bf u}_n,{\bf u}_{n+1})=\epsilon [1+\cos(\phi_n)], \label{f3}\\
\cos(\phi_n)=-({\bf v}_{n-1},{\bf v}_n)/r_{n-1}r_n, \nonumber
\end{eqnarray}
describes the deformation of the $n$th angle of the chain $\phi_n$, the energy $\epsilon =3.5$~eV defines the bending stiffness of the chain.

The third potential describes weak non-valent interactions of distant chain nodes $n$ and $k$.
These interactions can be described with high accuracy by the Lennard-Jones potential $(m,n)$
\begin{equation}
W_i(r)=\varepsilon_i[m(\rho_i/r)^n-n(\rho_i/r)^m]/(n-m), \label{f4}
\end{equation}
with powers $m=5$, $n=11$, interaction energy $\varepsilon_1=0.00832$~eV and equilibrium length $\rho_1=3.607$~\AA~ \cite{Savin2019prb} (index $i=1,2,3$ specifies the potential number).
The potential $W_1({\bf u}_{n},{\bf u}_k)=W_1(r_{nk})$, where the distance between the nodes of the chain $r_{nk}=|{\bf u}_{n}-{\bf u}_k|$.
The parameters of the potentials (\ref{f2}) and (\ref{f3}) were determined in \cite{Savin2015prb,Savin2015ftt} from the analysis of dispersion curves of graphene nanoribbons.

Consider a nanoribbon lying on a flat polymer substrate -- on the surface of a crystal of polyethylene
(PE) (CH$_2)_\infty$. 
Let us assume that the zigzag chains of PE move like a solid and always lie across the nanoribbon parallel to its surface (Fig. \ref{fig01}).
Then each PE molecule can be replaced by an effective united atom located in the $xz$ plane.

The Hamiltonian (\ref{f1}) defines the energy of the nanoribbon falling on its longitudinal band of width $\Delta_y=3r_c/2=2.127$~\AA. 
The polyethylene macromolecule has the shape of a flat zigzag with pitch $r_{pe}=1.53$~\AA~ (the length of the valence bond CH$_2$--CH$_2$) and angle $\alpha=110^\circ$.
The longitudinal zigzag pitch is $a_y=r_{pe}\sin(\alpha/2)$, the zigzag width is $a_x=r_{pe}\cos(\alpha/2)$.
The mass of the atoms of the zigzag lying in longitudinal part of the length $\Delta_y$ is $M_{pe}=14m_p \Delta_y/a_y=23.76m_p$.

In the $xz$ plane, a particle of mass $M_{pe}$ corresponds to each PE macromolecule.
The interaction of the particles (averaged over shifts and rotations  interaction of the trans-zigzag PE molecules) is described with a high accuracy by the Lennard-Jones potential (\ref{f4}) $W_2(r)$ with powers $m=5.5$, $n=11$ and equilibrium distance $\rho_2=4.62$~\AA.
The length of the zigzag $\Delta_y$ accounts for the interaction energy $\varepsilon_2=0.0324$~eV \cite{Savin2021vms}.

The polymer substrate of the nanoribbon is modeled by a 2D lattice of particles of mass $M_{pe}$ with the Hamiltonian
\begin{eqnarray}
H_2=\sum_{n=N_c+1}^{N}\frac12M_{pe}(\dot{\bf u}_n,\dot{\bf u}_n)+Z({\bf u}_n)\nonumber\\
+\sum_{n=N_c+1}^{N-1}\sum_{k=n+1}^N W_2({\bf u}_n,{\bf u}_{k}), \label{f5}
\end{eqnarray}
where the total number of particles $N=N_c+N_{p}$, $N_p$ is the number of particles of the polymer substrate (number of PE molecules).
In the 2D model, each node of the lattice having coordinates ${\bf u}_n=(x_n,z_n)$ corresponds to the position of the projection of the center of the $n$th zigzag PE chain on the $xz$ plane -- see Fig. \ref{fig01} (b).

The first term of the Hamiltonian (\ref{f5}) defines the kinetic energy of the lattice.
The potential $Z({\bf u})$ describes the interaction of the lattice nodes (PE macromolecules) with the solid flat substrate $z\le 0$ on which the lattice lies. 
The interaction energy of the particles with the half-space $z\le 0$ is described by the Lennard-Jones potential (3,9) \cite{Zhang2013,Zhang2014,Aitken2010}:
\begin{equation}
Z({\bf u})=Z(z)=\varepsilon_0 [(h_0/z)^9-3(h_0/z)^3]/2, \label{f6}
\end{equation}
where $\varepsilon_0$ is the interaction energy (adhesion energy), $h_0$ is the equilibrium distance to the surface. 
For the surface of a silicon oxide crystal SiO$_2$ energy $\varepsilon_0=0.075$~eV, distance $h_0=5$~\AA~ \cite{Koenig2011}.
The potential $W_2({\bf u}_n,{\bf u}_{k})=W_2(r_{nk})$ defines the interaction of the polymer substrate particles.

The interaction of the nanoribbon (chain nodes) with the polymer substrate is given by the sum
\begin{equation}
E_{int}=\sum_{n=1}^{N_c}\sum_{k=N_c+1}^N W_3({\bf u}_n,{\bf u}_{k}), \label{f7}
\end{equation}
where the interaction potential of the chain nodes with the polymer substrate particles $W_3({\bf u}_n,{\bf u}_{k})=W_3(r_{nk})$ is also described by the Lennard-Jones potential (\ref{f4}) with powers $m=5.5$, $n=11$, energy $\varepsilon_3=0.0162$~eV, and equilibrium distance $\rho_3=4.125$~\AA~ \cite{Savin2021vms}.

The 2D model of a multilayer graphene sheet lying on a soft polymer substrate is schematically represented in Fig. \ref{fig01} (c). 
Here chains of small particles correspond to graphene layers, large black particles correspond to macromolecules of the soft polymer substrate, the black straight line shows the fixed surface of the SiO$_2$ crystal on which the polymer substrate lies. The small arrows show the action of the external pressure $p$.
Along the $x$-axis, periodic boundary condition is used.

Potential energy of the system
\begin{equation}
E=E_1+E_2+E_3+E_4, \label{f8}
\end{equation}
where the energy of the $K$-layer nanoribbon is
\begin{eqnarray}
E_1=\sum_{k=1}^K\sum_{n=1}^{N_c}[V({\bf u}_{n_k},{\bf u}_{n_k+1})
+U({\bf u}_{n_k-1},{\bf u}_{n_k},{\bf u}_{n_k+1})
\nonumber\\
+\frac12\sum_{|l-n|>5}W_1({\bf u}_{n_k},{\bf u}_{l_k})],~~n_k=n+(k-1)N_c,~~
\label{f9}
\end{eqnarray}
interaction energy of the nanoribbon layers is
\begin{equation}
E_2=\sum_{k=1}^{K-1}\sum_{l=k+1}^K\sum_{n=1}^{N_c}\sum_{m=1}^{N_c} W_1({\bf u}_{n_k},{\bf u}_{m_l}),
\label{f10}
\end{equation}
the energy of the polymer substrate is
\begin{equation}
E_3=\sum_{n=N_K+1}^{N}Z({\bf u}_n)
+\sum_{n=N_K+1}^{N-1}\sum_{k=n+1}^N W_2({\bf u}_n,{\bf u}_{k}), \label{f11}
\end{equation}
where the number of chain links is $N_K=KN_c$, the total number of particles is $N=N_K+N_p$,
interaction energy of the nanoribbon with the polymer substrate is
\begin{equation}
E_4=\sum_{n=1}^{N_K}\sum_{k=N_K+1}^N W_3({\bf u}_n,{\bf u}_{k}). \label{f12}
\end{equation}
Here, the vector ${\bf u}_{n_k}$, for $n_k=(k-1)N_c+n$, $n=1,...,N_c$, $k=1,...,K$, specifies the position of the $n$-th node of $k$-th chain, and the vector ${\bf u}_{N_K+l}$, $l=1,...,N_p$, defines the position of $l$-th particle of the polymer substrate. The particle mass is
$M_n=M_c$ for $n\le N_K$ and $M_n=M_{pe}$ for $n>N_K$.

Let us take a polymer substrate consisting of $N_y$ dense layers of $N_x$ particles ($N_p=N_x\times N_y$) lying on a rigid substrate $z\le 0$.
The period of such a lattice along the $x$ axis is $L_x=N_xb$, where $b=4.5519$~\AA~ is the distance between nearest neighbors in a two-dimensional crystal of polymer particles.
Let us cover this polymer lattice with a $K$-layer chain (nanoribbon) of $N_c=L_x/a$ links. 
In the initial state the particles of this system will have coordinates
\begin{eqnarray}
&&x_n=(s_k+l-1)a_0,~z_n=L_z+h_1(k-1),\nonumber\\
&&n=(k-1)N_c+l,~k=1,...,K,~l=1,...,N_c,\label{f13}\\
&&x_n=(s_j+i-1)b,~z_n=h_0+(j-1)h_2,\nonumber\\
&&n=KN_c+(j-1)N_x+i,~i=1,...N_x,~j=1,...,N_y, \nonumber
\end{eqnarray}
where $s_k=0$ for odd and $s_k=1/2$ for even $k$, $a_0=L_x/N_c$, $h_1=3.33$~\AA, $h_2=b\sqrt{3}/2$,
$L_z=h_0+N_yh_2$.

To model the effect of pressure on our two-component molecular system, we introduce an external force ${\bf f}$, which is applied to each node of the external chain (chain number $k=K$) orthogonally to its line. 
The amplitude of the force $f=|{\bf f}|$ determines the magnitude of the pressure $p=f/a\Delta_y=f/\sqrt{3}a^2$. If the chain forms a line orthogonal to the $z$-axis in the ground state, then the external force ${\bf f}=(0,-a^2p\sqrt{3})$.

To find the basic homogeneous state of the system we must to solve the problem for the minimum of potential energy
\begin{equation}
E=E_1+E_2+E_3+E_4+E_5\rightarrow\min~:~\{{\bf u}_n\}_{n=1}^N, \label{f14}
\end{equation}
where the last term is responsible for the pressure action on the outer surface of the nanoribbon,
\begin{equation}
E_5=\sum_{n=(K-1)N_c+1}^{KN_c} a^2p\sqrt{3}z_n. \label{f15}
\end{equation}
The minimum problem (\ref{f14}) was solved numerically using the conjugate gradient method.
The initial configuration of the molecular system (\ref{f13}) was used to find the ground state of the molecular system.
\begin{figure}[tb]
\begin{center}
\includegraphics[angle=0, width=1.0\linewidth]{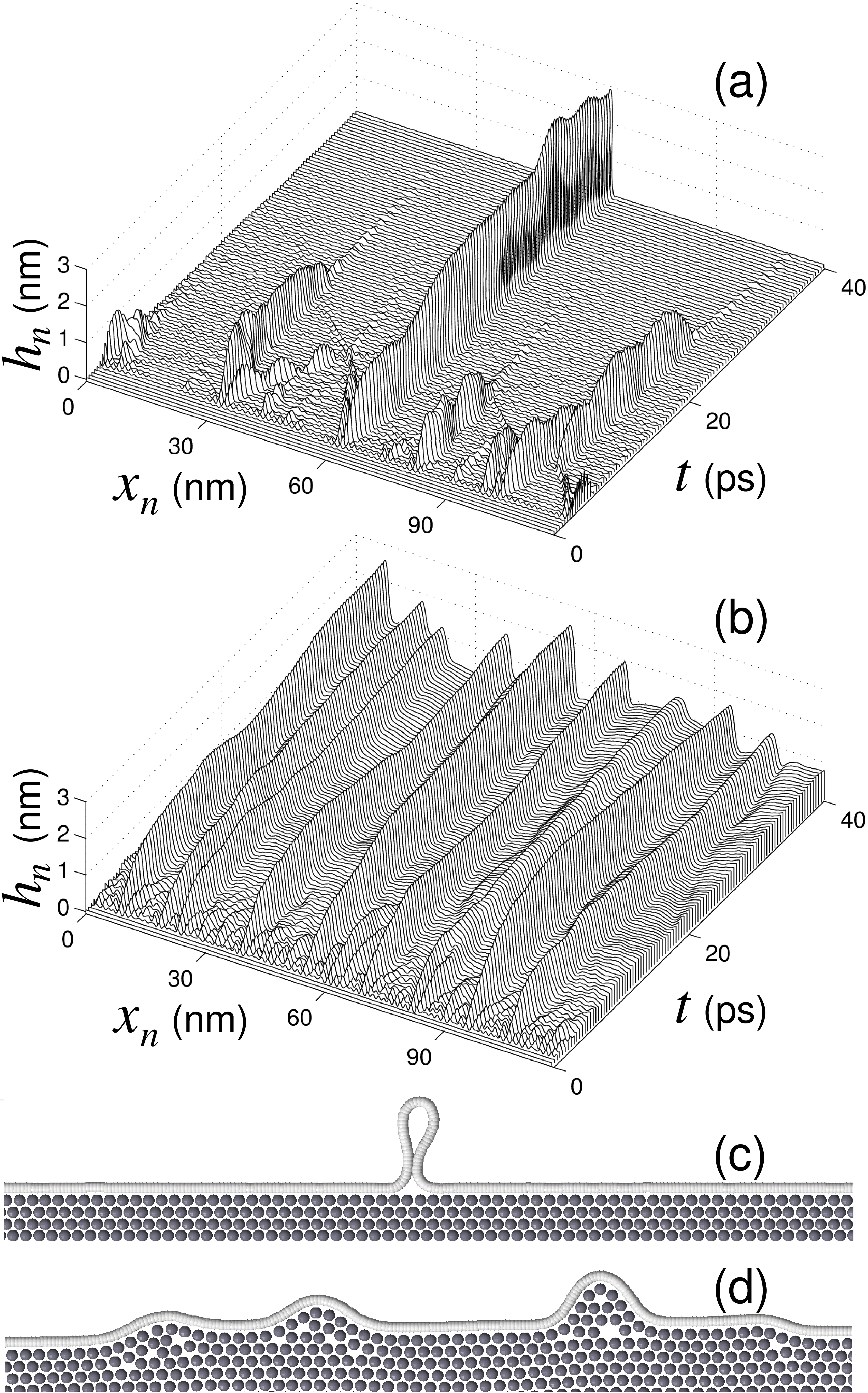}
\end{center}
\caption{\label{fig02}\protect
The formation in a uniformly compressed molecular system of a single-layer nanoribbon lying on a polymer substrate of thickness $L_z=20.15$~nm under compression $d=0.05$ of (a) a vertical fold at a fixed position of substrate atoms, (b) a wrinkle system when the mobility of substrate atoms is taken into account.
The number of chain links is $N_c=1005$, the number of substrate atoms is $N_p=271\times 50$ (period $L_x=117.19$~nm), temperature is $T=1$K, external pressure is $p=150.6$~bar. 
The dependence on time $t$ of the nanoribbon shape $\{ x_n,h_n=z_n-h_0\}_{n=1}^{N_c}$ is shown.
Parts (c) and (d) show the shape of the nanoribbon with the substrate at a finite moment of time.
}
\end{figure}

\section{Critical value of longitudinal compression}

Let us take the solution of the minimum problem (\ref{f13}) $\{ {\bf u}_n^0=(x_n^0,z_n^0)\}_{n=1}^N$ and model its homogeneous compression along the $x$-axis.
For this purpose, we numerically integrate the system of Langevin equations of motion
\begin{eqnarray}
M_n\ddot{\bf u}_n&=&-\frac{\partial H}{\partial {\bf u}_n}-\Gamma M_n\dot{\bf u}_n-\Xi_n-f{\bf w}_n/2a,\label{f16}\\
&& (K-1)N_c< n\le KN_c, \nonumber \\
M_n\ddot{\bf u}_n&=&-\frac{\partial H}{\partial {\bf u}_n}-\Gamma M_n\dot{\bf u}_n-\Xi_n,\label{f17}\\
&& 1\le n\le (K-1)N_c,~n>KN_c, \nonumber
\end{eqnarray}
with the initial condition
\begin{equation}
{\bf u}_n(0)=\left((1-d)x_n^0,z_n^0\right),~~\dot{\bf u}_n(0)=0, \label{f18}
\end{equation}
and period along the $x$ axis $(1-d)L_x$, where $d\in [0,1)$ is the longitudinal compression ratio (compression percentage is $d\cdot 100$\%). 
Here, equation (\ref{f16}) specifies the motion of the top chain (the top sheet of graphene), which is acted upon by an external pressure $p=f/\sqrt{3}a^2$, vector ${\bf w}_n=(z_{n+1}-z_{n-1}, x_{n-1}-x_{n+1})$ defines the direction orthogonal to the chain at node $n$ (normalized orthogonal vector ${\bf e}_n={\bf w}_n/|{\bf w}_n|\simeq {\bf w}_n/2a$).
The friction coefficient is $\Gamma=1/t_r$, the relaxation time is $t_r=10$~ps, $\Xi_n=(\xi_{n,1},\xi_{n,2})$ is a two-dimensional vector of normally distributed random Langevin forces with correlation functions
$$
\langle\xi_{n,i}(t_1)\xi_{k,j}(t_2)\rangle=2M_nk_BT\Gamma\delta_{nk}\delta_{ij}\delta(t_2-t_1)
$$
($k_B$ is Boltzmann constant, $T$ is thermostat temperature).

The equations of motion Eq. (\ref{f16}), (\ref{f17}) are solved numerically using the velocity Verlet method \cite{Swope1982}.
A time step of 1 fs is used in the simulations since further reduction of the time step has no appreciable effect on the results.

Let us first consider the approximation of a non-deformable substrate, when only atoms of chains (graphene sheet) with numbers $1\le n\le KN_c$ ($K$ is the number of layers of the sheet) can participate in the motion.
This approximation is usually used when modeling the formation of wrinkles and folds in a graphene sheet located on a flat surface of a solid crystal.
Quasi-analytical models based on calculus of variations \cite{Zhu2012,Zhang2013a,Box2015,Aljedani2020,Aljedani2020a,Cox2020,Aljedani2021}, models based on continuum mechanics using the finite element method \cite{Zhang2013,Zhang2014} and full-atom models using molecular dynamics \cite{Mulla2015,Zhu2020,Zhao2020} were used to describe individual wrinkles and folds. 

A detailed analysis of wrinkle and fold formation in graphene sheets lying on a non-deformable flat substrate was carried out in \cite{Savin2024ftt}.
Uniaxial compression of such a sheet leads to the formation of a localized convex wrinkle with an empty bubble-like region between the sheet and the substrate. 
With increasing compression, the wrinkles fold (collapse) and form vertically standing folds (spikes) with dense multilayer foots and drop-shaped heads.
When the sheet slides freely on the substrate, the interaction of wrinkles and folds is reduced to tugging of the part of the sheet located between them. 
As a consequence, the interaction of two wrinkles leads to the enlargement of the larger wrinkle at the expense of the disappearance of the smaller one. 
The interaction of two folds can only lead to a change in their shape. 
For this reason, a small uniaxial compression may produce only one wrinkle in a sheet, while a strong compression may produce several stable vertical folds.

Numerical analysis of the system of equations of motion (\ref{f16}), (\ref{f17}) has shown that the flat edge of a fixed regular two-dimensional lattice of substrate particles allows almost free sliding of the chain. 
For example, for a sample of size $123.36\times 20.23$~nm$^2$ ($N_p=271\times 50$, $p=0$), sliding of the whole chain of $N_c=1005$ links on the fixed two-dimensional lattice requires overcoming the energy barrier $\Delta E<0.097$~eV.
Therefore, the above-described scenario of wrinkle and fold formation will be fulfilled here.

Numerical integration of the system of equations of motion (\ref{f16}), (\ref{f17}) for particles with $1\le n\le KN_c$ fully confirmed this scenario.
Here, a homogeneously compressed flat nanoribbon with $d=0.05$ first produces a small-amplitude periodic ripples from which a system of localized wrinkles is formed. 
Then the largest wrinkle grows due to the absorption of small ones. 
The growth of this wrinkle leads to its collapse and to the formation of a vertical fold -- see Fig. \ref{fig02} (a,c).
This scenario is also realized for multilayer nanoribbons and at high values of external pressure $p<2000$~bar.
The number of vertical folds formed depends on the length of the nanoribbon $L_x$ and on the magnitude of its compression $d$ (the larger $L_x$ and $d$, the larger the number of vertical folds formed).

The type of nanoribon dynamics changes dramatically if we take into account the mobility of the soft polymer substrate. For this let us integrate the system of equations of motion (\ref{f16}), (\ref{f17}) for all particles (for $1\le n\le N=KN_c+N_p$).
Here, the longitudinal compression of the two-component molecular system nanoribbon/polymer (NR/PE) results only in leads to the formation on the nanoribbon of a system of non-interacting wrinkles, whose interior is filled with polymer substrate particles -- see Fig. \ref{fig02} (b,d).
\begin{figure}[tb]
\begin{center}
\includegraphics[angle=0, width=1.0\linewidth]{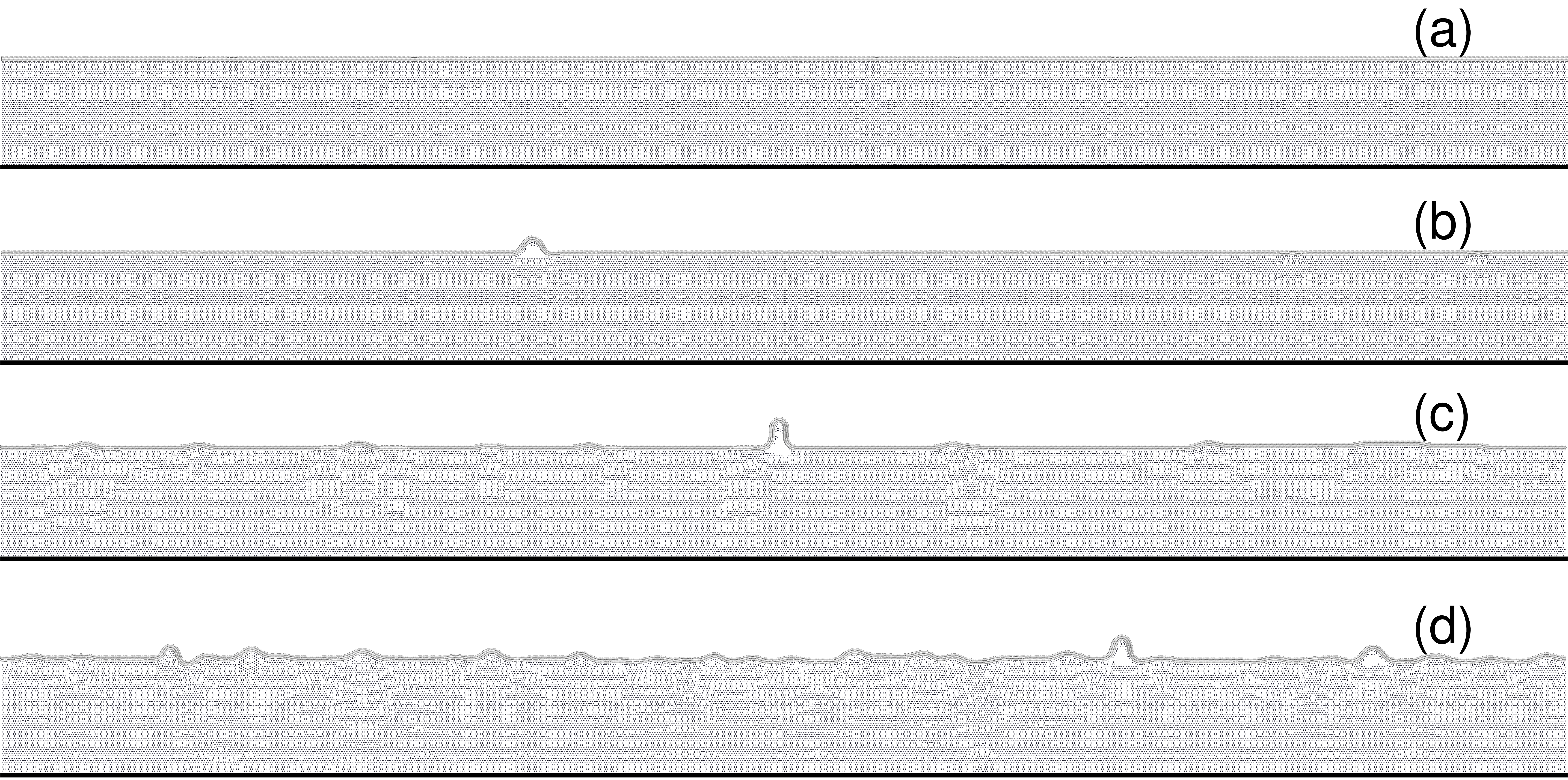}
\end{center}
\caption{\label{fig03}\protect
Structures formed by uniform compression of the NR/PE molecular system of a three-layer nanoribbon ($K=3$) lying on a polymer substrate of thickness $L_z=20.23$~nm under compression of (a) $d=0.009$, (b) 0.010, (c) 0.030, and (d) 0.050.
The periodic cell of system (period $L_x=(1-d)299.97$~nm) at time $t=0.57$~ns is shown(external pressure $p=150$ bar, temperature $T=1$~K).
}
\end{figure}
\begin{figure}[tb]
\begin{center}
\includegraphics[angle=0, width=1.0\linewidth]{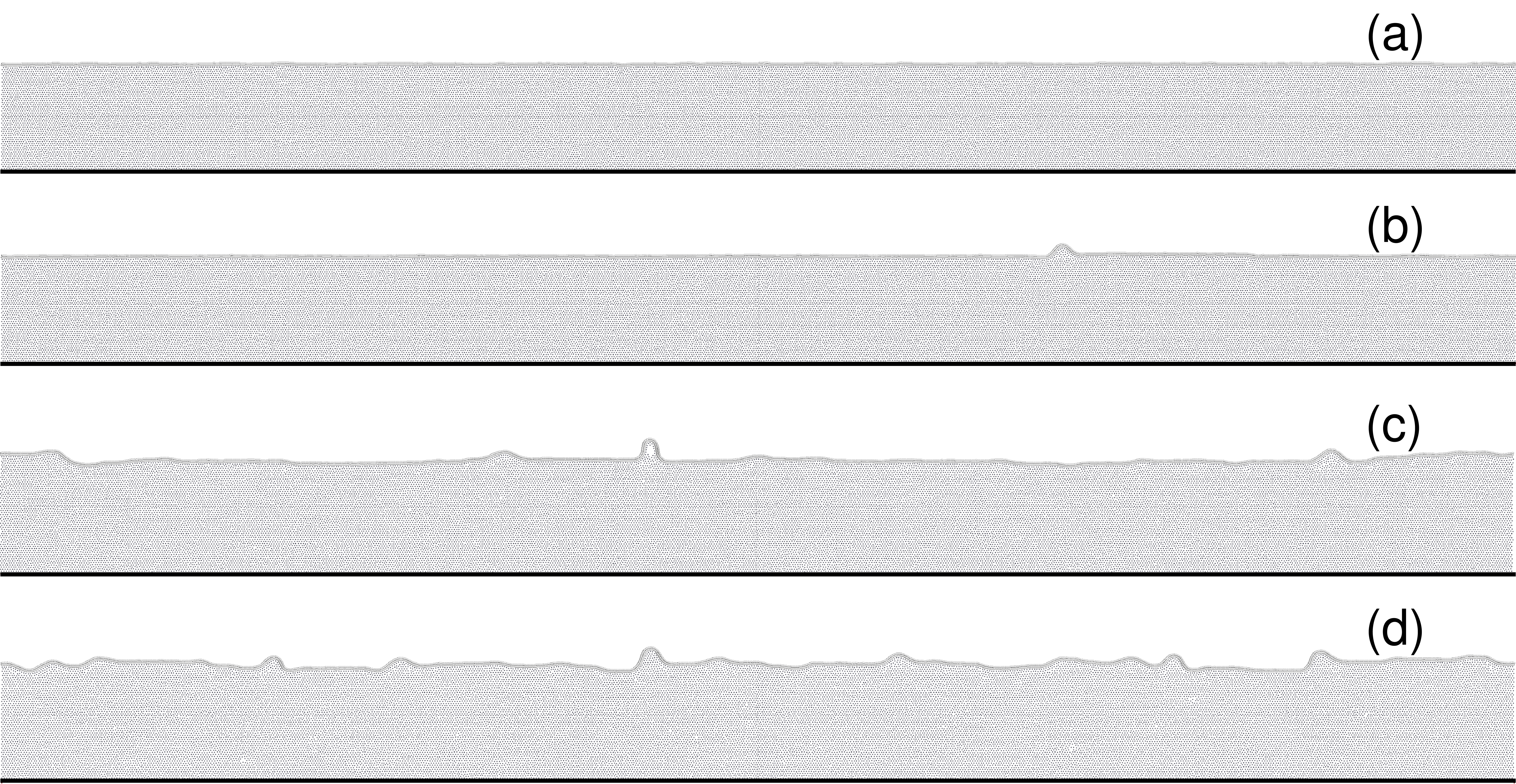}
\end{center}
\caption{\label{fig04}\protect
Structures formed by uniform compression of the NR/PE molecular system of a two-layer nanoribbon ($K=2$) lying on a polymer substrate of thickness $L_z=20.23$~nm under compression of (a) $d=0.007$, (b) 0.008, (c) 0.030, and (d) 0.050.
The periodic cell of system (period $L_x=(1-d)299.97$~nm) at time $t=0.74$~ns is shown (external pressure $p=150$ bar, temperature $T=100$~K).
}
\end{figure}

Let us determine the critical value of longitudinal compression $d_0$ at which the basic homogeneous state of the NR/PE system loses stability.

The solution of the problem (\ref{f14}) allows us to obtain the basic homogeneous state of the molecular system $\{ {\bf u}^0_n=(x_n^0,z_n^0)\}_{n=1}^N$ at an external pressure $p\ge 0$.
To check the stability of this state against longitudinal compression, we numerically integrate the system of equations of motion (\ref{f16}), (\ref{f17}) with initial condition (\ref{f18}) at temperature $T=1$~K. 
We take $N_x=659$, $N_y=50$, then the periodic cell of the polymer substrate will consist of $N_p=N_x\times N_y=32950$ particles, period is $L_x=299.97$~nm, number of atoms in one chain is $N_c=2444$. 
A  visualization of the ground homogeneous state of the two-component molecular NR/PE system is shown in Fig. \ref{fig03} (a).
\begin{table}[tb]
\caption{
Dependence of the critical value of compression $d_0$ on pressure $p$ for a $K$-layer sheet of graphene lying on a polymer substrate at temperature $T=1$ and 100~K.
\label{tab1}
}
\begin{tabular}{cc|cccccc}
 $T$ (K)& ~~$K$~~ & ~~$p=0$~~ & ~15~ & ~~150~~ & ~~1500~~ & 15000&(bar) \\
 \hline
 ~  & 1 & 0.018 & 0.018 & 0.018 & 0.020 & 0.030&\\
 1 & 2 & 0.012 & 0.012 & 0.012 & 0.014 & 0.021&\\
 ~  & 3 & 0.010 & 0.010 & 0.010 & 0.011 & 0.017&\\
 \hline
  ~  & 1 & 0.009 & 0.009 & 0.009 & 0.014 & 0.024&\\
100 & 2 & 0.008 & 0.008 & 0.008 & 0.011 & 0.019&\\
  ~  & 3 & 0.007 & 0.007 & 0.007 & 0.009 & 0.016&\\
\hline
\end{tabular}
\end{table}

Numerical integration of the system of equations of motion has shown that at each value of pressure $p$ there is a critical value of longitudinal compression $d_0$, at which there is a loss of stability of the homogeneous state: this state remains stable at compression $d<d_0$ and collapses at $d>d_0$.
Dependence of the critical value $d_0$ on the external pressure $p$, the number of graphene layers $K$ at temperature $T=1$~K and 100~K is presented in table \ref{tab1}.
As can be seen from the table, the value of $d_0$ increases monotonically with increasing $p$.
The high pressure $p\ge 150$~bar leads to a noticeable additional stabilization of the homogeneous state.
On the other hand, thermal fluctuations lead to a significant decrease of the critical value, particularly noticeable at low pressures.

The structures arising from the uniform compression of a molecular system are shown in Figs. \ref{fig03} and \ref{fig04}.
At pre-critical compression $d<d_0$, the homogeneous state remains stable during the whole time of numerical simulation. 
When $d=d_0$, the graphene sheet forms a single wrinkle, and when $d>d_0$ it forms a system of wrinkles. 
Note that even at very low temperature $T=1$~K (Fig. \ref{fig03}) the area under the wrinkle is partially filled with molecules of the polymer substrate, and at higher temperatures -- complete filling occurs (Fig. \ref{fig04}). 
This filling prevents the interaction of the wrinkles and the formation of from its vertical folds (spikes).
\begin{figure}[tb]
\begin{center}
\includegraphics[angle=0, width=1.0\linewidth]{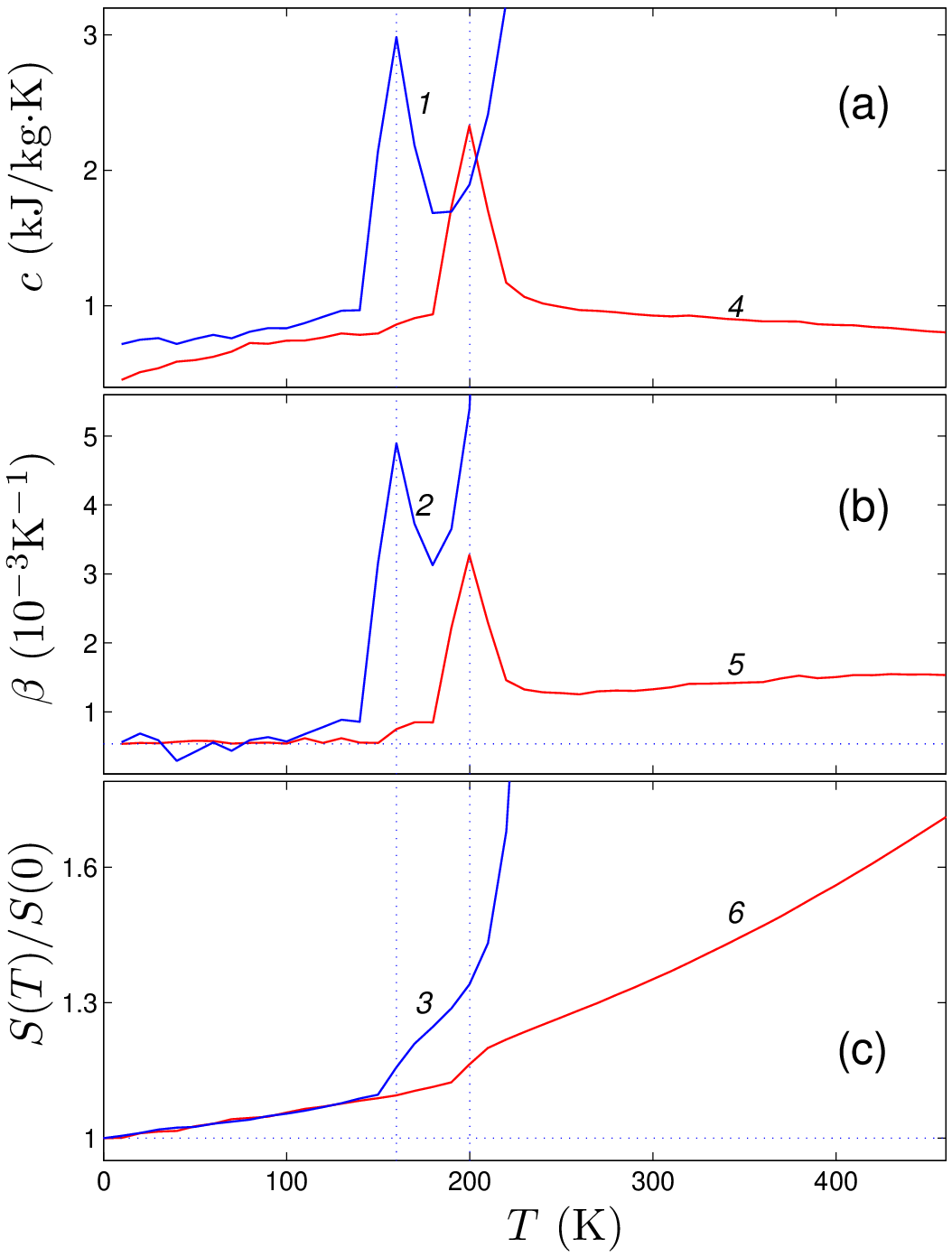}
\end{center}
\caption{\label{fig05}\protect
Dependence on temperature $T$ of (a) the heat capacity of the NR/PE system $c$, (b) the coefficient of volumetric thermal expansion $\beta$, and (c) the dimensionless volume $S(T)/S(0)$. 
Curves 1, 2, 3 and 4, 5, 6 show the dependencies at external pressures $p=150$ and 1500~bar.
The vertical dotted lines show the temperature values $T=160$ and 200~K.
}
\end{figure}
\begin{figure}[tb]
\begin{center}
\includegraphics[angle=0, width=1.0\linewidth]{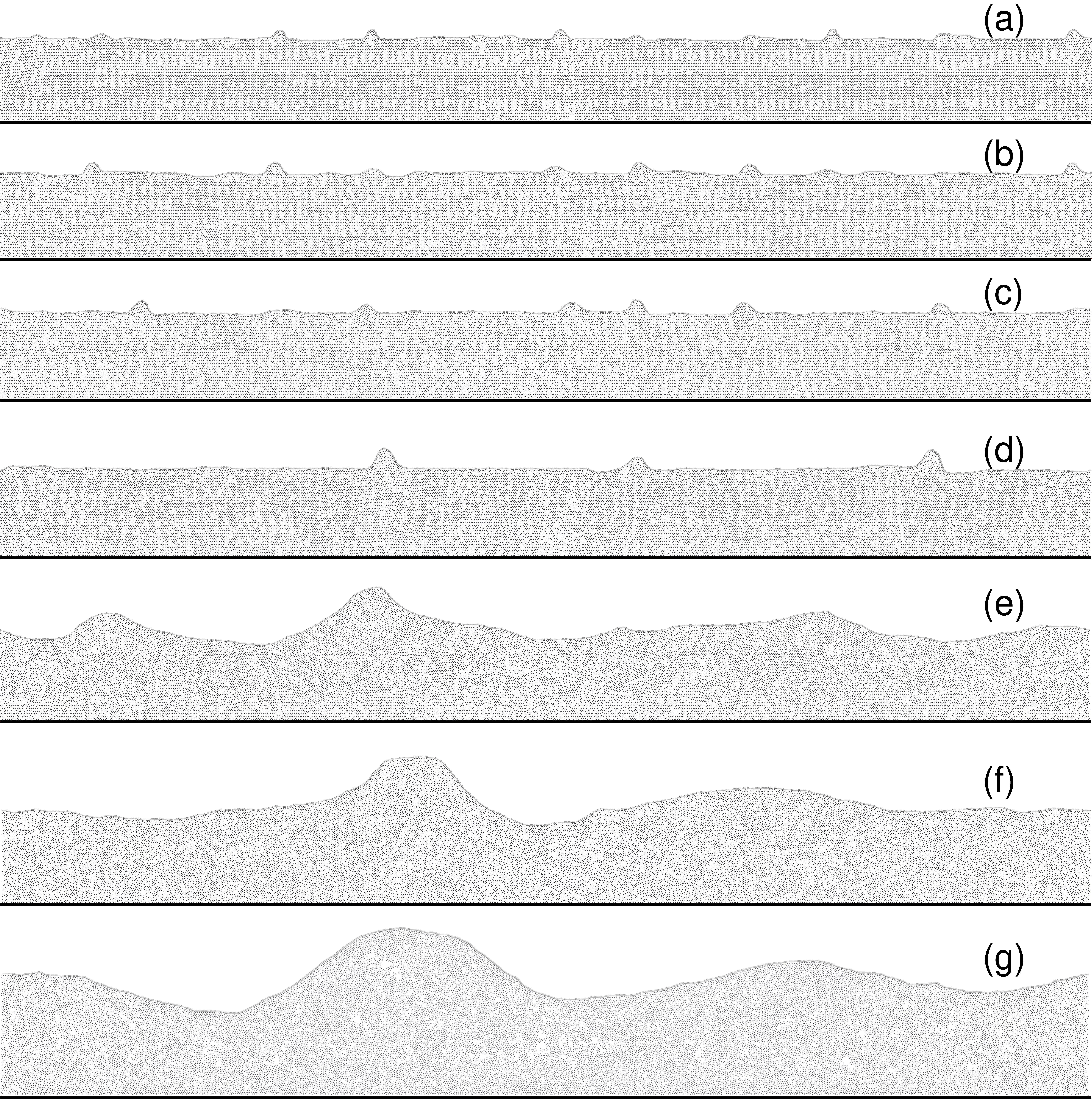}
\end{center}
\caption{\label{fig06}\protect
Structures formed by uniform compression of the NR/PE molecular system from a single-layer nanoribbon ($K=1$) lying on a polymer substrate of $N_p=659\times 50$ particles at temperatures of (a) $T=50$, (b) 100, (c) 130, (d) 150, (e) 160, (f) 180, and (g) 200~K.
Periodic cell (period $L_x=(1-d)299.97$~nm, $N_c=2444$) at time $t=10$~ns is shown, external pressure $p=150$ bar, compression $d=0.05$.
}
\end{figure}
\begin{figure}[tb]
\begin{center}
\includegraphics[angle=0, width=1.0\linewidth]{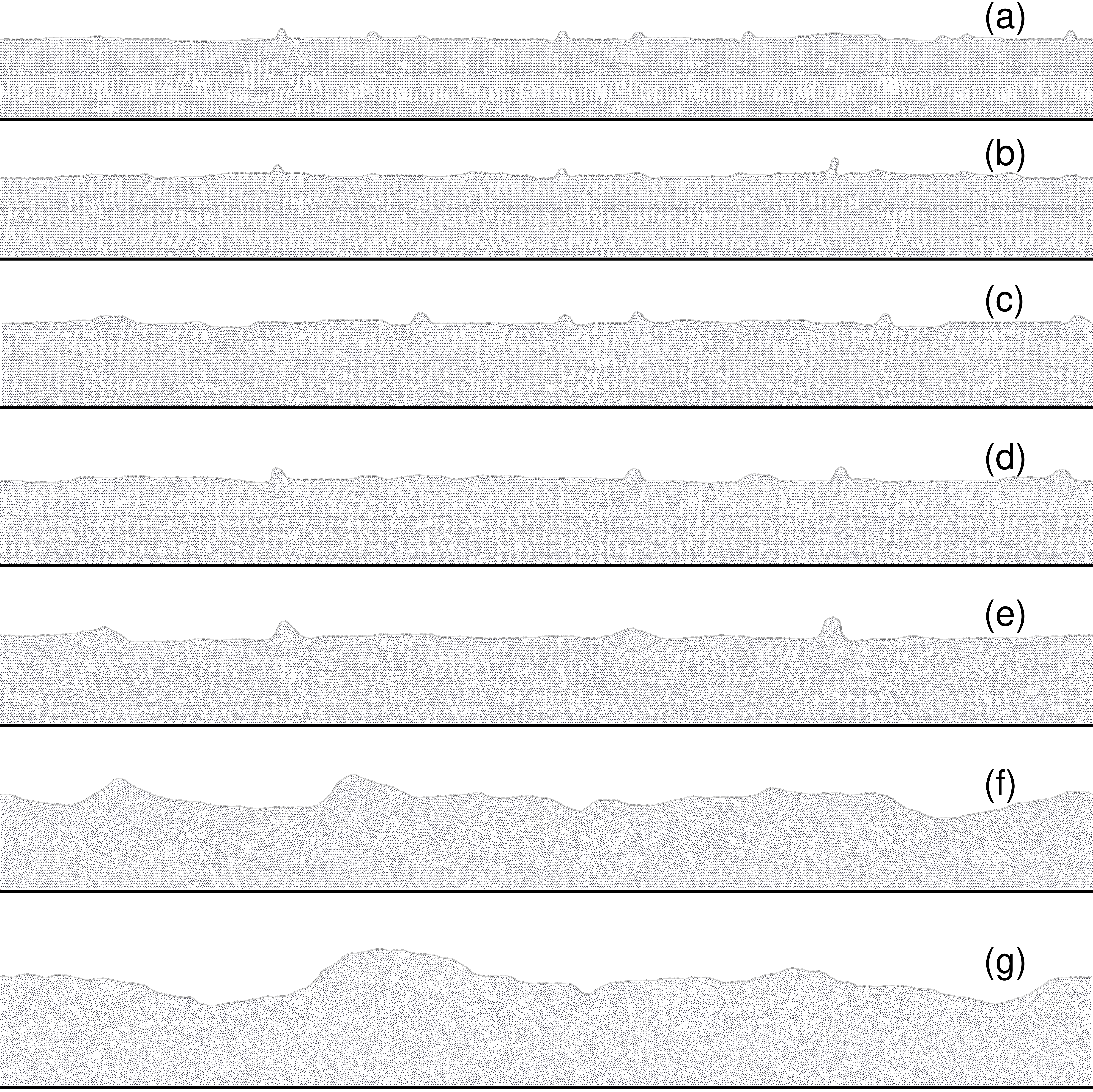}
\end{center}
\caption{\label{fig07}\protect
Structures formed by uniform compression of the NR/PE molecular system from a single-layer nanoribbon ($K=1$) lying on a polymer substrate of $N_p=659\times 50$ particles at temperatures of (a) $T=50$, (b) 100, (c) 150, (d) 170, (e) 190, (f) 200, and (g) 300~K.
Periodic cell at time $t=10$~ns is shown, external pressure $p=1500$ bar, compression $d=0.05$.
}
\end{figure}

\section{Influence of temperature}

Let us study the influence of thermal fluctuations on the structural changes of a two-component molecular system NR/PE under its homogeneous compression. 
For this purpose, we consider the dynamics of a periodic cell consisting of $N_p=659\times 50$ particles of polymer substrate (period $L_x=299.97$~nm, number of atoms in one chain $N_c=2444$) under compression $d=0.05$. 
For this purpose, we will numerically integrate the system of Langevin equations (\ref{f16}), (\ref{f17}) with initial condition (\ref{f18}) and period $(1-d)L_x$ at different values of thermostat temperature $T$.

The system of equations of motion was integrated numerically over $t=10$~ns. 
After obtaining equilibrium state of molecular system with the thermostat, we have found the average values of the energy $\bar{E}(T)$ and the the area (volume) occupied by the polymer substrate under the chain $\bar{S}(T)$,
$$
S=\frac12\sum_{n=1}^{N_c} z_n(x_{n+1}-x_{n-1}),
$$
where the periodic boundary conditions are taken into account in the summation. 
Further, we found dependencies on temperature of heat capacity
$$
c(T)=\frac{1}{M}\frac{d}{dT} \bar{E},~~M=\sum_{n=1}^N M_n,
$$
and coefficient of thermal volumetric expansion
$$
\beta(T)=\frac{1}{\bar{S}}\frac{d}{dT} {\bar{S}}
$$
The change in the volume of the polymer substrate is conveniently described by its relative values $s(T)=S(T)/S(0)$.

The dependence of the values $c$, $\beta$, $s$ on the temperature $T$ is shown in Fig. \ref{fig05}.
As can be seen from the figure, there are two characteristic temperature values: $T_1<T_2$.
At the first temperature $T_1$, melting by the polymer substrate occurs, and a sharp increase in heat capacity and coefficient of volume expansion is noted. 
Here the polymer substrate goes from a 2D crystal state to a liquid state. 
At the second temperature $T_2$, continuous volume growth occurs -- there is a gradual transition of the liquid state to the gaseous state.
The values of these temperatures increase with increasing external pressure, so at $p=0$~bar $T_1=155$~K ($T_2=180$), at $p=150$ $T_1=160$ ($T_2=200$), and at $p=1500$ $T_1=200$ ($T_2>480$).
Note that these temperature values do not depend on the number of chain (sheet) layers $K=1$, 2, 3, but are completely determined by the properties of the polymer substrate.

Note that the used simplified 2D model of crystalline polyethylene does not allow us to obtain accurate melting point values.
For medium- and high-density polyethylene the melting point is typically in the range
120 to 130$^\circ$C (393 to 403~K). 
Nevertheless, the 2D model allows us to qualitatively describe the melting behavior of the polymer substrate.

Structural changes of the two-component molecular system NR/PE with increasing temperature are shown in Figs. \ref{fig06} and \ref{fig07}.
\begin{figure}[tb]
\begin{center}
\includegraphics[angle=0, width=1.0\linewidth]{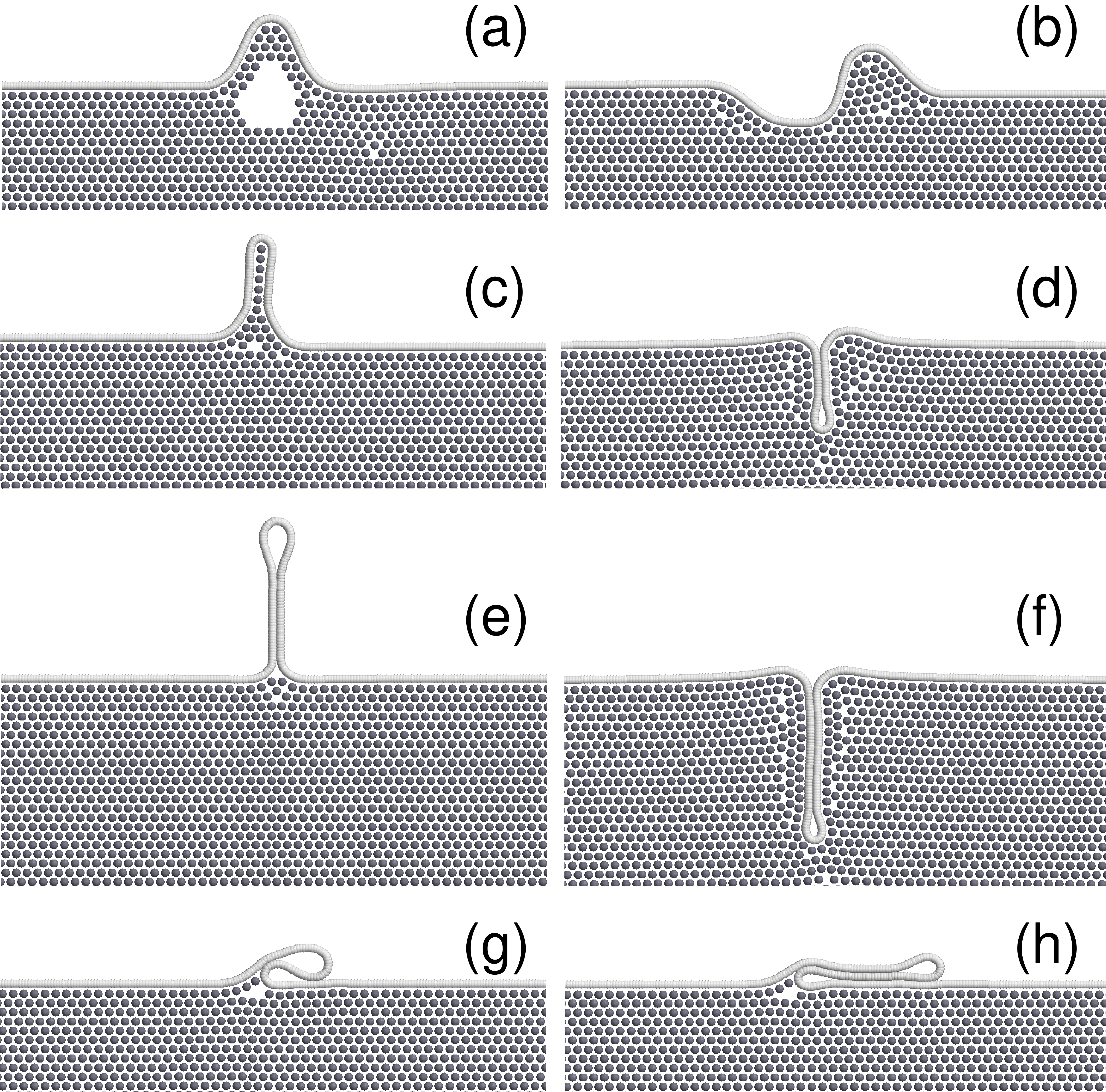}
\end{center}
\caption{\label{fig08}\protect
Folds of graphene sheet on soft polymer substrate at excess number of chain links (a, b) $\Delta N_c=30$, (c, d) 60, (e, f) 110. 
Parts a, c, e show vertical folds directed away from the substrate, parts b, d, f show vertical folds directed toward the interior of the substrate. Parts g and h show three-layer folds lying along the substrate surface at $\Delta N_c=60$ and 110.
}
\end{figure}

At low pressure $p=150$~bar, melting of the polymer substrate occurs at $T_1=160$~K.
At $T<160$K 5\% compression of the NR/PE system leads to the appearance of a set of small wrinkles in the covering sheet of graphene, the interior of which is completely filled with particles of the polymer substrate -- see Fig. \ref{fig06} (a,b).
An increase in temperature leads to an increase in the size of the wrinkles and to a decrease in their number (c), the wrinkles become largest near melting point (d).
Melting leads to the complete disappearance of the wrinkle system, the graphene sheet obtains a longitudinal wave-like structure lying on the liquid substrate (e,f).
Increasing temperature at $T>T_2=180$~K leads to a continuous increase in the volume of the substrate due to the formation of empty regions (lacunas) in it (g).

At stronger pressure $p=1500$~bar, the substrate melting occurs at $T_1=200$~K.
Here, at $T<200$~K 5\% compression leads to the appearance of the covering sheet of graphene system of small wrinkles, the interior of which is completely filled with substrate particles -- see Fig. \ref{fig07} (a,b,c,d). 
An increase in temperature leads to a decrease the number of wrinkles and the size of the wrinkles increases. Wrinkles become the largest near the melting point (e). 
Melting of the substrate leads to complete disappearance of the wrinkle system, the sheet of graphene takes the form of random long waves (f,g). 
At $200<T<480$~K, the substrate always remains in a liquid dense state.
Here, high pressure prevents the formation of empty regions in it.

Note that cooling of the molten substrate leads to its crystallization, but the system of of wrinkles on the surface is not restored. 
We can conclude that the melting of the polymer substrate of the compressed two-component graphene/polymer system leads to irreversible disappearance of the wrinkle system on the surface of the graphene sheet.
\begin{table}[tb]
\caption{
Dependence of the energy of stationary state of the graphene/polymer system on the number of nodes $\Delta N_c$ participating in the formation of a vertical fold directed upward ($E_\perp$), downward ($E_\top$) and of the surface fold ($E_=$).
\label{tab2}
}
\begin{tabular}{c|ccccccc}
 $\Delta N_c$ & 30 & 40 & 50 & 60 &70 & 90& 110 \\
 \hline
 ~$E_\perp$~(eV)~ & ~0.668 & ~0.812 &  ~1.072 & ~-0.452 &  ~1.004 &  ~0.395 & -0.110 \\
 $E_\top$~(eV)  & ~0.529 & ~0.276 & -0.422 & ~-0.978 & -1.847 & -3.450 & -4.921 \\
 $E_=$~(eV)     &   --  &  --   &  2.037 &  1.504 &  0.618 &  0.011 & -1.182 \\
\hline
\end{tabular}
\end{table}

\section{Fold changes during substrate melting}

Earlier we considered folds appearing on the surface of a two-component graphene/polymer system at its longitudinal compression. 
Let us take another situation when the length of the graphene nanoribbon is longer than the length of the polymer substrate, i.e. when $N_c>N_c^0=L_x/a$. 
This situation can occur when the graphene sheet is placed on the prepared polymer substrate. 
The excessive length of the sheet $\Delta L_c=(N_c-N_c^0)a>0$ (excessive number of chain links $\Delta N_c=N_c-N_c^0>0$) will lead to the appearance of its folds. 
The stationary fold can be obtained as a numerical solution of the minimum energy of the system (\ref{f14}).
In the case of rigid substrate, the fold of the sheet will always be directed outward (upward) from the substrate. 
For soft substrate, the fold can also be directed into the substrate (downward), entering deeply into it.

Consider for simplicity a single-layer sheet of graphene (number of layers $K=1$) at zero pressure ($p=0$).
Let us take a substrate consisting of $N_p=659\times 50$ particles, then the substrate period $L_x=299.97$~nm, and the number of chain atoms uniformly covering the whole substrate without folds $N_c^0=2444$.
The solution of the problem (\ref{f14}) at the number of chain links $N_c=N_c^0$ allows us to obtain the ground state of the graphene/polymer system. 
We will use the energy of this state $E_0=-3967.36$~eV as the energy reference level.

The view of possible folds of a graphene sheet on a polymer substrate at different values of the number of nodes $\Delta N_c$ participating in their formation are shown in Fig.~\ref{fig08}.
As can be seen from the figure, the upwardly directed vertical folds have a partial filling of the interior space with substrate particles. 
Vertical folds directed downward at $\Delta N_c<40$ have the form of of one period of a periodic wave tightly adjoining the substrate. 
At $\Delta N_c\ge 40$, the fold vertically enters the substrate as a bilayer of graphene.
Folds lying on the substrate can exist only at $\Delta N_c\ge 50$.
A three-layer section of the chain forms in the localization area.

The dependence of the fold energy on $\Delta N_c$ for their three possible types is provided in Table \ref{tab2}.
As can be seen from the table, the most favorable in energy are vertical folds directed into the substrate, since for them all particles of the chain always participate in the interaction with the substrate.
The folds lying on the substrate (three-layer chain sections) can exist only at $\Delta N_c\ge 50$.
They become more energetically favorable compared to upward-facing vertical fold at $\Delta N_c\ge 70$.
\begin{figure}[tb]
\begin{center}
\includegraphics[angle=0, width=1.0\linewidth]{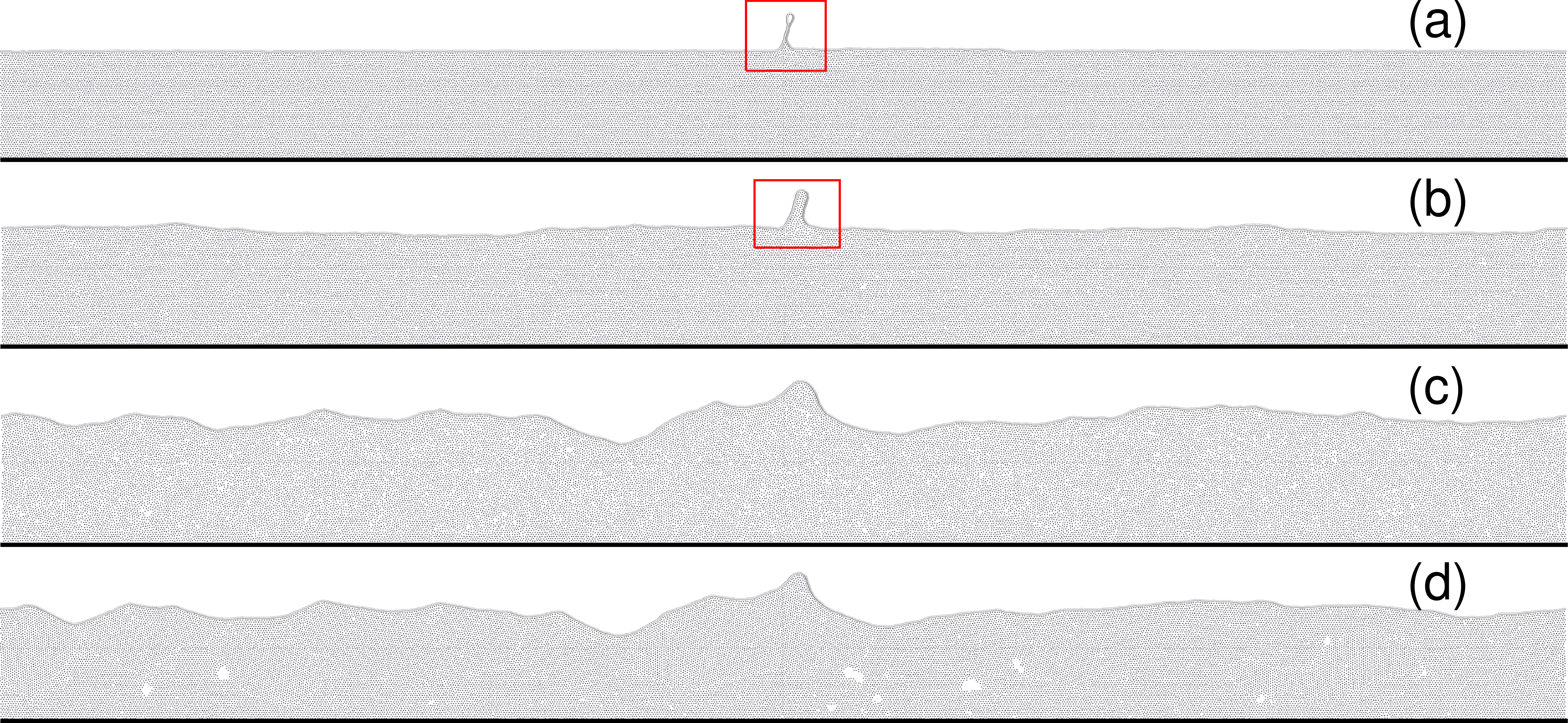}
\end{center}
\caption{\label{fig09}\protect
Structure of the NR/PE molecular system of a single-layer nanoribbon ($K=1$) lying on a polymer substrate of $N_p=659\times 50$ particles with excess number of nodes in the chain $\Delta N_c=110$ at temperature (a) $T=100$~K (vertically standing bilayer fold), (b) 150~K (vertically standing fold filled with substrate particles), (c) 160~K (disappearance of the fold due to melting of the substrate) and (c) when the system is further cooled to 100~K (nanoribbon lying on the undulating surface of the polymer substrate). The region of NR/PE system with fold is highlighted by the red rectangle.
}
\end{figure}

\begin{figure}[tb]
\begin{center}
\includegraphics[angle=0, width=1.0\linewidth]{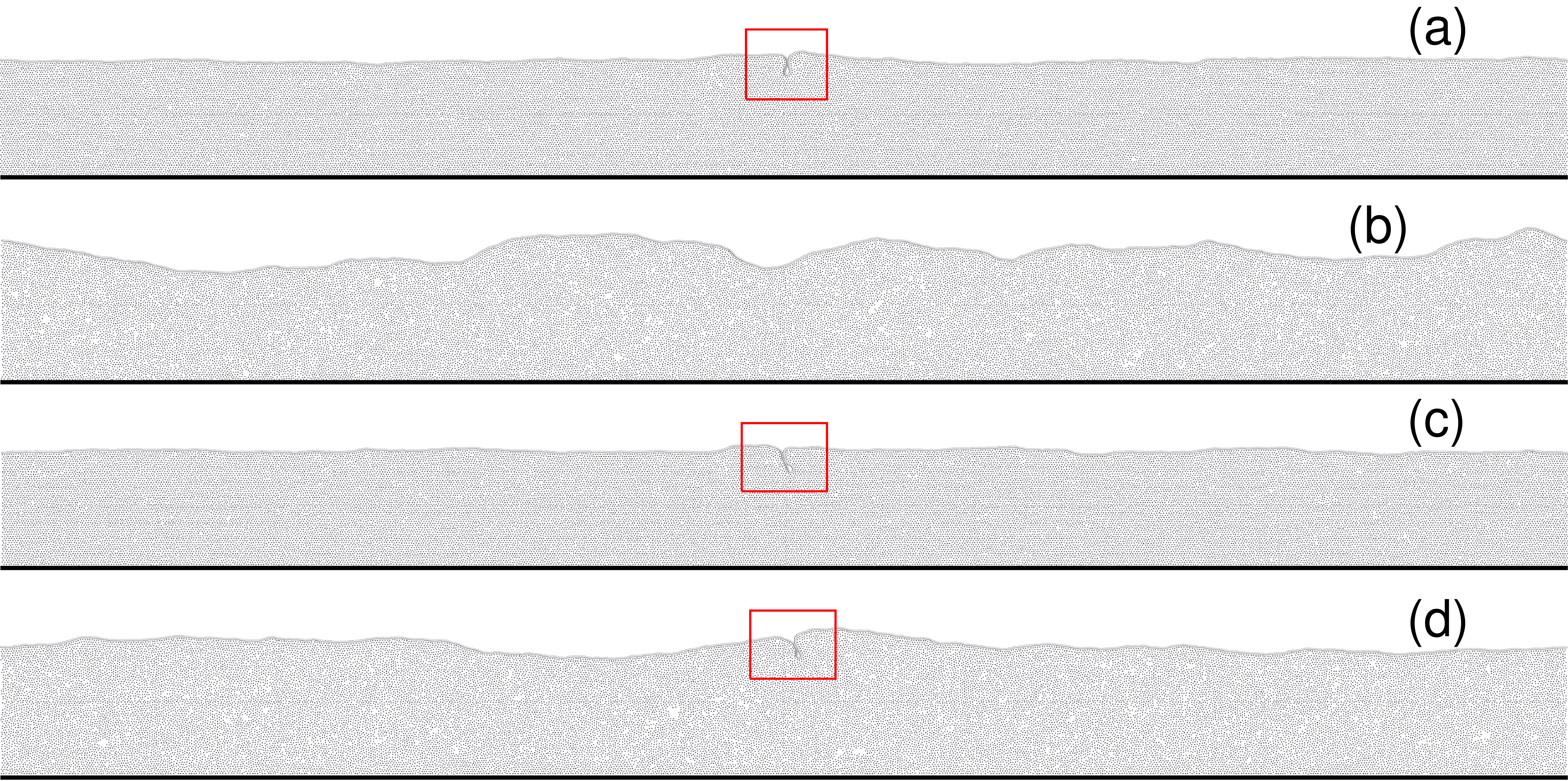}
\end{center}
\caption{\label{fig10}\protect
Structure of the NR/PE molecular system of a single-layer nanoribbon ($K=1$) lying on a polymer substrate of $N_p=659\times 50$ particles with excess number of nodes in the chain $\Delta N_c=70$ at temperature (a) $T=150$~K (vertical fold directed into the substrate), (b) 170~K (disappearance of the fold as a result of substrate melting) and at (c) $\Delta N_c=90$, $T=150$~K and (d) 170~K (vertical fold in the molten substrate). 
The region of NR/PE system with fold is highlighted by the red rectangle.
}
\end{figure}

\begin{figure}[tb]
\begin{center}
\includegraphics[angle=0, width=1.0\linewidth]{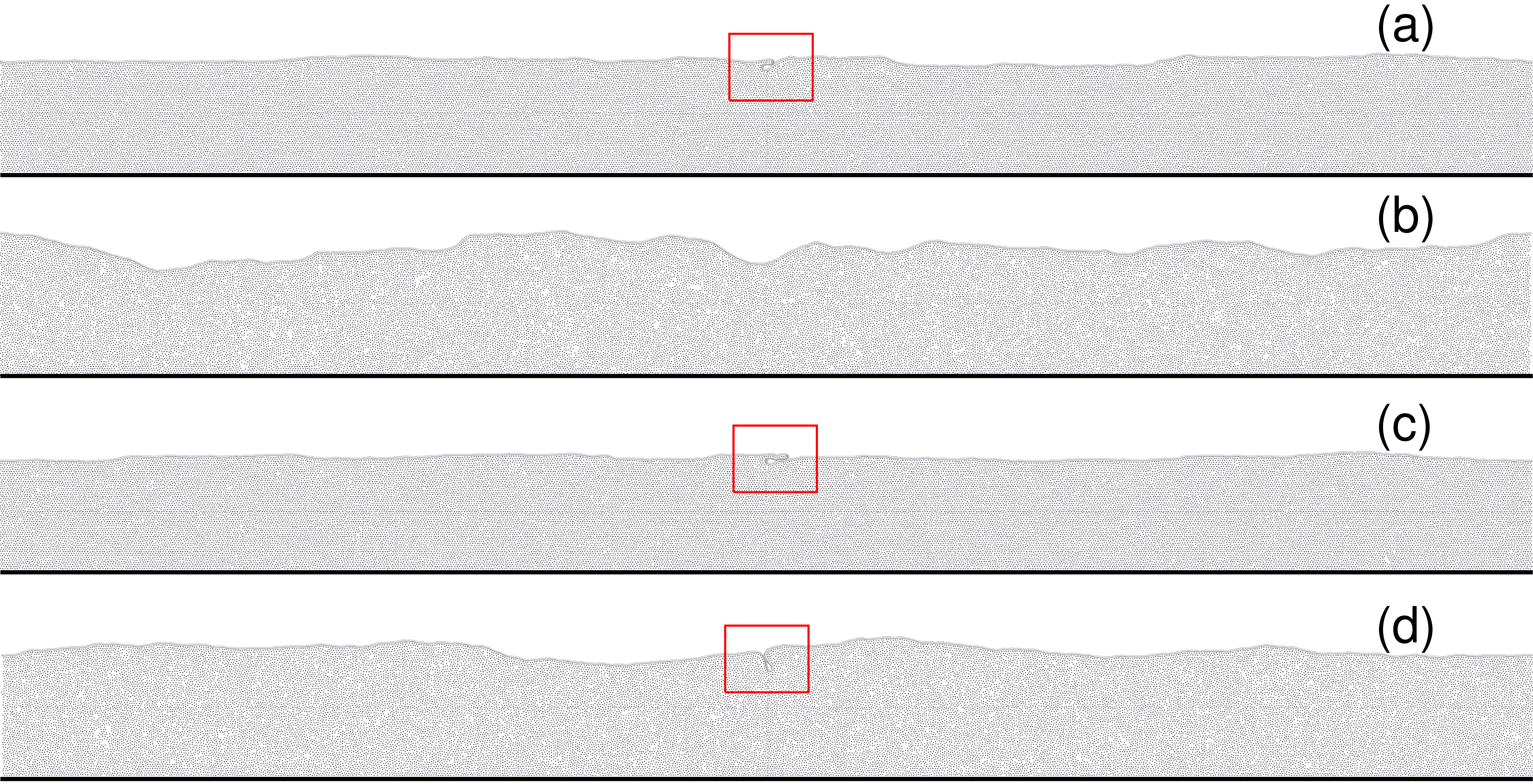}
\end{center}
\caption{\label{fig11}\protect
Structure of NR/PE molecular system consisting of a single-layer nanoribbon ($K=1$) lying on a polymer substrate of $N_p=659\times 50$ particles with excess number of nodes in the chain $\Delta N_c=70$ at temperature (a) $T=150$~K (lying three-layer fold), (b) 170~K (disappearance of the fold due to the melting of the substrate) and at (c) $\Delta N_c=90$, $T=150$~K and (d) 170~K (transition of the three-layer chain section into a fold directed into the molten substrate).
The region of NR/PE system with fold is highlighted by the red rectangle.
}
\end{figure}

Thermal vibrations of a soft substrate can lead to high-amplitude deformations.
Let us check how these deformations affect the folded structures of graphene sheets.
For this purpose, we consider the dynamics of a periodic NR/PE cell with the number of chain nodes $N_c>N_c^0$ at different values of temperature. 
Numerical integration of the system of Langevin equations (\ref{f16}), (\ref{f17}) with the initial condition corresponding to the stationary state of the fold has shown that at temperature $T$ below the melting temperature $T_1$ the structure of the folds does not change (their shape may only change slightly) -- see Fig. \ref{fig09}, \ref{fig10}, and \ref{fig11}.
The transition of the substrate to the liquid state at $T=T_1=155$~K leads to a significant change in the shape of the chain. The surface of the molten substrate takes the form of an irregular wave -- see Fig. \ref{fig09} (c). As a result, its surface area increases, which leads to stretching of the chain and, consequently, to the reduction of its fragments involved in the formation of the folds.

The modeling has shown that at an excessive number of chain links $\Delta N_c\le 110$, melting of the substrate always leads to the disappearance of vertical folds directed outside of the substrate. 
As a result, the chain takes the shape of a curve at each point adjacent to the substrate surface -- see Fig. \ref{fig09}.
This fold-free chain shape is retained if we lower the temperature by converting the substrate to a solid state.
Here, melting of the substrate allows us to get rid of the folds completely.

Vertical folds directed inside the substrate are more resistant to melting.
These folds disappear completely only at $\Delta N_c <80$ -- see Fig. \ref{fig10} (a,b).
At $\Delta N_c>80$, melting of the substrate does not lead to complete disappearance of the folds.
Although the folds decrease in size, they persist and remain begin directed inside the liquid substrate -- see Fig. \ref{fig10} (c,d).

The folds lying on the substrate surface and forming three-layer chain sections can exist only at $\Delta N_c\ge 50$. 
Melting of the substrate leads to the disappearance of these folds only at $\Delta N_c<80$. At $\Delta N_c>80$, the three-layer chain section sinks into the liquid substrate, forming a more energetically favorable inward-directed vertical fold -- see Fig. \ref{fig11}. 
Note that when overlapping a graphene sheet on a solid (unmelted) flat substrate, only the folds which are directed vertical upward and are lying along the surface can be formed.
For formation of vertical folds directed inside the substrate it is necessary to melt it. 
Then the folds lying on the surface with $\Delta N_c>80$ will sink into the substrate and turn into vertical folds directed inside.
When the temperature is lowered and the substrate solidifies, they will retain their shape.

Thus, in order to obtain inward-directed vertical folds, it is necessary to melt the substrate. On the other hand, melting of the substrate leads to the disappearance of all small folds with excess number of links $\Delta N_c< 80$.

\section{Conclusion}

The modeling has shown that the formation of wrinkles and folds in a sheet of graphene (single-layer or multilayer) located on a soft polymer substrate has a number of essential features.
Unlike flat surfaces of rigid crystals, particles of soft deformable polymer substrates can penetrate into wrinkles and folds of graphene sheet filling the voids between the sheet and the substrate.

For uniaxial compression of the two-component graphene/polymer system there is a critical value of compression $d_0$ at which the initial homogeneous state of the system loses stability.
High external pressure $p\ge 150$~bar leads to a noticeable additional stabilization of the homogeneous state, i.e. to an increase of $d_0$. 
On the other hand, an increase in temperature $T$ leads to a significant decrease of the critical value, especially noticeable at low pressure values.
At uniaxial compression above the critical value the graphene sheet has a system of non-interacting localized wrinkles, the interior of which is filled with molecules of the polymer substrate.
This filling prevents the interaction of wrinkles and the formation of large vertical folds (spikes).
An increase in temperature leads to an increase in the size of wrinkles and to a decrease in their number.
Wrinkles become the largest before starting of the substrate melts.
Melting leads to the complete disappearance of localized wrinkles, and the graphene sheet takes a longitudinal wave-like structure lying on the liquid substrate.

The excessive length of the graphene sheet when it is placed on a flat polymer substrate can lead to the creation of three types of stable localized folds: vertical folds directed upward from the substrate; folds lying on the substrate (in the area of their localization a three-layer sheet area is formed); vertical folds directed into the substrate and penetrating it.
The last type of folds cannot exist for rigid crystal surfaces.
For a soft deformable substrate, the most favorable in terms of energy are folds directed into the substrate, since for them the interaction between the sheet and the substrate is most fully realized.

Melting of the substrate leads to the disappearance of all wrinkles and small folds.
Cooling of the melted substrate leads to its crystallization, but the wrinkle system of the graphene sheet on the surface is not restored.
Therefore, melting of the substrate and its subsequent cooling can serve as a method to get rid of localized wrinkles and folds of the graphene sheet.

\end{document}